\documentclass[prd,preprintnumbers,showpacs,onecolumn,showkeys,floatfix,nofootinbib,notitlepage]{revtex4-1}
\usepackage{graphicx}
\usepackage{amssymb}
\usepackage{amsmath}
\usepackage{mathtools}
\usepackage{epsfig}
\usepackage{color}
\usepackage{enumerate}
\usepackage{slashed}
\usepackage{hyperref}
\usepackage{epstopdf}

\def\slashchar#1{\setbox0=\hbox{$#1$}
   \dimen0=\wd0 \setbox1=\hbox{/} \dimen1=\wd1
   \ifdim\dimen0\big>\dimen1 \rlap{\hbox to \dimen0{\hfil/\hfil}} #1
   \else  \rlap{\hbox to \dimen1{\hfil$#1$\hfil}} / \fi}

\newcommand{\ud}{\mathrm{d}}
\newcommand{\be}{\begin{equation}}
\newcommand{\ee}{\end{equation}}
\newcommand{\bea}{\begin{eqnarray}}
\newcommand{\eea}{\end{eqnarray}}

\newcommand{\Appendix}[1]%
    {%
     \section{#1}%
      }

\begin{document}

\title{The Continuity of the Gauge Fixing Condition $n\cdot\partial n\cdot A=0$ for $SU(2)$ Gauge Theory}

\author{Gao-Liang Zhou}
\email{ zhougl@itp.ac.cn}
\affiliation{College of Science, Xi'an University of Science and Technology, Xi'an 710054, People's Republic of China}
\author{Zheng-Xin Yan}
\affiliation{College of Science, Xi'an University of Science and Technology, Xi'an 710054, People's Republic of China}
\author{Xin Zhang}
\affiliation{College of Science, Xi'an University of Science and Technology, Xi'an 710054, People's Republic of China}




\begin{abstract}
The continuity of the gauge fixing condition $n\cdot\partial n\cdot A=0$ for $SU(2)$ gauge theory on the manifold $R\bigotimes S^{1}\bigotimes S^{1}\bigotimes S^{1}$ is studied here, where $n^{\mu}$ stands for directional vector along $x_{i}$-axis($i=1,2,3$).  It is proved that the gauge fixing condition is continuous given that gauge potentials are differentiable with continuous derivatives on the manifold $R\bigotimes S^{1}\bigotimes S^{1}\bigotimes S^{1}$ which is compact.
\end{abstract}


\maketitle

\section{Introduction.}
\label{introduction}

It is well known that the Faddeev-Popov quantization\cite{Faddeev:1967fc} of non-Abelian gauge theory is hampered by the Gribov ambiguity \cite{Gribov-1978,Singer-1978}. Especially, as proved in  \cite{Singer-1978},  there is no continuous gauge condition that is free from the Gribov ambiguity for non-Abelian gauge theories  on $3$-sphere($S^{3}$) and $4-$sphere($S^{4}$) once the gauge group is compact.  The conventional Faddeev-Popov quantization procedure\cite{Faddeev:1967fc} of gauge theory is based on the equation:
\begin{equation}
\label{gaugefix}
\int[\mathcal{D}\alpha(x)]\textrm{det} (\frac{\delta(G(A))}{\delta \alpha})\delta(G(A))=1,
\end{equation}
where $\alpha(x)$ is the parameter of Gauge transformation, $G(A)$ represents the gauge fixing function. According to results in \cite{Singer-1978}, it is impossible to choose suitable continuous gauge fixing function $G(A)$ so that the equation (\ref{gaugefix}) holds for non-Abelian gauge theory  on $3$-sphere($S^{3}$) and $4-$sphere($S^{4}$) given that the gauge group is compact. In addition, degeneracy of the gauge fixing condition relies on configurations of gauge potentials and may affect calculations of physical quantities.

While concerning infinitesimal gauge transformations, the Gribov ambiguity originates from zero eigenvalues(with nontrivial eigenvectors) of the Faddeev-Popov operator\cite{Gribov-1978,vanBaal:1991zw,Sobreiro:2004us,Vandersickel:2012tz}. Thus one can work in the region in which the Faddeev-Popov operator is positive definite to eliminate infinitesimal Gribov copies. Such region is termed as Gribov region in literatures\cite{Gribov-1978,Vandersickel:2012tz}. Studies on the Gribov region are interesting and fruitful(see, e.g. Refs.\cite{Zwanziger:1982na,Dell'Antonio:1991xt,Zwanziger:1988jt,Zwanziger:1989mf,Zwanziger:1992qr,Dudal:2007cw,
Dudal:2008sp,Capri:2012wx,Su:2014rma,Bandyopadhyay:2015wua,Guimaraes:2015vra}). The Gribov region method is extended to linear covariant gauges in \cite{Lavrov:2013boa} through the field dependent BRST transformation\cite{Joglekar:1994tq,Lavrov:2013rla,Moshin:2016xjq}. It can also be extended to other gauge conditions(see, e.g. Refs. \cite{Capri:2006cz,Capri:2010an,Capri:2015pfa,Capri:2016aif,Capri:2016aqq}). We emphasize that the Gribov ambiguity does not vanish even for one works in the Gribov region, although expectation values of gauge invariant quantise are not affected by such ambiguity while working  in the Gribov region\cite{Zwanziger:2003cf}. Other works on the Gribov ambiguity can been seen in  \cite{Serreau:2012cg,Serreau:2013ila,Pereira:2013aza,Pereira:2014apa,Bassetto:1983rq} and references therein.

In \cite{zhou:2016gb}, We present a new gauge condition $n\cdot \partial n\cdot A=0$ for non-Abelian gauge theory on the manifold $R\otimes S^{1}\otimes S^{1}\otimes S^{1}$, where $n^{\mu}$ is the directional vector along $x_{i}$-axis($i=1,2,3$). We have proved that $n\cdot \partial n\cdot A=0$ is a continuous gauge for non-Abelian gauge theory on $R\otimes S^{1}\otimes S^{1}\otimes S^{1}$ given that  generators of 
Wilson lines along $n^{\mu}$ are continuous on the manifold $R^{4}$, where the generator of a unitary matrix $U(x)=\exp(i\phi(x))$ means the Hermitian matrix $\phi(x)$ not infinitesimal generators of Lie group in this paper.  In addition, we have proved that the gauge condition is free from the Gribov ambiguity  except for configurations with zero measure.

In this paper, we study $SU(2)$ gauge theory on $R\otimes S^{1}\otimes S^{1}\otimes S^{1}$ to investigate the continuity of the gauge condition $n\cdot \partial n\cdot A=0$. Although the theory is quite simple, studies on $SU(2)$ gauge theory are helpful for understanding properties of the Gribov ambiguity of non-Abelian gauge theories. The work in this paper is also meaningful for the study on effects of the Gribov ambiguity on the weak interaction.    We will prove that the gauge fixing condition $n\cdot \partial n\cdot A=0$ is continuous for the theory considered here once gauge potentials are differentiable with continuous derivatives.   To simplify the proof, we consider the theory on the manifold $R\otimes S^{1}\otimes S^{1}\otimes S^{1}$ with finite lengths along every direction(including the time direction) in this paper. That is to say, the manifold considered here is compact. For the case that the length along the time direction tends to $\infty$,  the conclusion is not affected given that gauge potentials satisfy suitable boundary conditions.

The paper is organized as follows. In Sec.\ref{R3torus}, we describe the gauge theory on $R\otimes S^{1}\otimes S^{1}\otimes S^{1}$ briefly. In Sec\ref{continuity}, we consider $SU(2)$ gauge theory on $R\otimes S^{1}\otimes S^{1}\otimes S^{1}$ and present the proof that the gauge condition $n\cdot \partial n\cdot A =0$ is continuous given that gauge potentials are differentiable with continuous derivatives on the manifold.  Our conclusions and some discussions are presented in Sec.\ref{conc}.

\section{Gauge Theories on $R\otimes S^{1}\otimes S^{1}\otimes S^{1}$}
\label{R3torus}

In this section, we describe the gauge theory on the finite $3+1$ dimensional surface $R\otimes S^{1}\otimes S^{1}\otimes S^{1}$.  The manifold considered here can be obtained from the manifold $R^{4}$ through the identification
\begin{equation}
(t,x_{1},x_{2},x_{3})\sim (t,x_{1}+L_{1},x_{2},x_{3})\sim (t,x_{1},x_{2}+L_{2},x_{3})\sim (t,x_{1},x_{2},x_{3}+L_{3}),
\end{equation}
where $L_{i}$($i=1,2,3$) are  large constants. In addition, we require that the surface is finite along the time direction. That is,
\begin{equation}
\label{timeupb}
-T\le t \le T,
\end{equation}
where $T$ is a large (positive)constant. It is clear that the manifold considered here is compact.
We consider continuous gauge potentials on the surface $R\otimes S^{1}\otimes S^{1}\otimes S^{1}$ in this paper. We have,
\begin{eqnarray}
\label{periocond}
A^{\mu}(t,x_{1}+L_{1},x_{2},x_{3})&=&A^{\mu}(t,x_{1},x_{2},x_{3})
\nonumber\\\
A^{\mu}(t,x_{1},x_{2}+L_{2},x_{3})&=&A^{\mu}(t,x_{1},x_{2},x_{3})
\nonumber\\
A^{\mu}(t,x_{1},x_{2},x_{3}+L_{3})&=&A^{\mu}(t,x_{1},x_{2},x_{3}).
\end{eqnarray}
Gauge potentials are continuous functions on the compact manifold. As a result, we have,
\begin{equation}
|A^{\mu a}(x)|<\infty,
\end{equation}
where $a$ represents the color index. For simplicity, we consider $SU(2)$ gauge theory in this paper and have $a=1,2,3$.
Effects of center vortexes like those shown in\cite{'tHooft:1977hy} are not considered here. As a result, we require that
\begin{eqnarray}
\label{perioga}
U(t,x_{1}+L_{1},x_{2},x_{3})&=&U(t,x_{1},x_{2},x_{3})
\nonumber\\\
U(t,x_{1},x_{2}+L_{2},x_{3})&=&U(t,x_{1},x_{2},x_{3})
\nonumber\\
U(t,x_{1},x_{2},x_{3}+L_{3})&=&U(t,x_{1},x_{2},x_{3}),
\end{eqnarray}
for continuous gauge transformation on the manifold considered here.

We study the continuity of the gauge condition $n\cdot \partial n\cdot A=0$ in this paper, where $n^{\mu}$ represents directional vectors along $x_{i}$-axis($i=1,2,3$). Without loss of generality, we take $n^{\mu}$ as
\begin{equation}
n^{\mu}=(0,0,0,1).
\end{equation}
It is proved in \cite{zhou:2016gb} that the gauge condition $n\cdot \partial n\cdot A=0$ is continuous on $R\otimes S^{1}\otimes S^{1}\otimes S^{1}$ given that generator of the Wilson line
\begin{equation}
\label{Wilsonline}
W(x_{0},x_{1},x_{2},x_{3})\equiv\mathcal{P}\exp (\mathrm{i}g\int_{0}^{x_{3}}\ud z  n\cdot A(x_{0},x_{1},x_{2},z))
\end{equation}
is a continuous on the manifold $R^{4}$, where the generator of a unitary matrix $U(x)=\exp(i\phi(x))$ should be understood  as the Hermitian matrix $\phi(x)$ not infinitesimal generators of Lie group in this paper. For Abelian gauge theory, generator of the Wilson line¡¡(\ref{Wilsonline}) is a continuous function on $R^{4}$ given that $n\cdot A(x)$ is continuous $R^{4}$. In fact, the Wilson line (\ref{Wilsonline}) can be written as
\begin{equation}
\label{Wison-abelian}
\exp (\mathrm{i}g\int_{0}^{x_{3}}\ud z  n\cdot A(x_{0},x_{1},x_{2},z))
\end{equation}
for Abelian gauge theory. Generator of the Wilson line (\ref{Wison-abelian}) reads
\begin{equation}
g\int_{0}^{x_{3}}\ud z  n\cdot A(x_{0},x_{1},x_{2},z),
\end{equation}
which is continuous on $R^{4}$. Thus the transformation
\begin{equation}
V(x)\equiv\exp (\mathrm{i}g\frac{x_{3}}{L_{3}}\int_{0}^{L_{3}}\ud z  n\cdot A(x_{0},x_{1},x_{2},z))
\exp (-\mathrm{i}g\int_{0}^{x_{3}}\ud z  n\cdot A(x_{0},x_{1},x_{2},z))
\end{equation}
is continuous on $R\otimes S^{1}\otimes S^{1}\otimes S^{1}$. One can verity that
\begin{equation}
n\cdot \partial n\cdot A^{V}=0
\end{equation}
for Abelian gauge theory, where
\begin{equation}
n\cdot A^{V}(x)=n\cdot A(x)+\frac{\mathrm{i}}{g}Vn\cdot\partial V^{\dag}(x).
\end{equation}

For non-Abelian  gauge theory, the situation is more subtle. For example, we consider the following Wilson line
\begin{equation}
\widetilde{W}(x)\equiv\left\{
\begin{array}{cc}
\exp( \mathrm{i}\pi\cos(\frac{2\pi x_{3}}{L_{3}})\sigma_{1})      &(x_{3}\le 0)
\\
\exp( \mathrm{i}\pi\cos(\frac{2\pi x_{3}}{L_{3}})\sigma_{3})      &(x_{3}\ge 0)
\end{array}
\right. ,
\end{equation}
where $\sigma_{i}$($i=1,2,3$) are Pauli matrixes. The gauge potential corresponding to such Wilson line  reads,
\begin{equation}
n\cdot A(x)=\left\{
\begin{array}{cc}
-\frac{2\pi^{2}\sigma_{1}}{gL_{3}}\sin (\frac{2\pi x_{3}}{L_{3}})
\exp( \mathrm{i}\pi\cos(\frac{2\pi x_{3}}{L_{3}})\frac{\sigma_{1}}{2})      &(x_{3}\le 0)
\\
-\frac{2\pi^{2}\sigma_{3}}{gL_{3}}\sin (\frac{2\pi x_{3}}{L_{3}})
\exp( \mathrm{i}\pi\cos(\frac{2\pi x_{3}}{L_{3}})\frac{\sigma_{3}}{2})      &(x_{3}\ge 0)
\end{array}
\right. ,
\end{equation}
which is continuous on $R\otimes S^{1}\otimes S^{1}\otimes S^{1}$. The generator of the Wilson line reads,
\begin{equation}
\widetilde{\phi}(x)\equiv\left\{
\begin{array}{cc}
(2N_{1}(x)+\cos(\frac{2\pi x_{3}}{L_{3}}))\sigma_{1}      &(x_{3}\le 0)
\\
 (2N_{2}(x)+\cos(\frac{2\pi x_{3}}{L_{3}}))\sigma_{1}     &(x_{3}\ge 0)
\end{array}
\right. ,
\end{equation}
where $N_{1}(x)$ and $N_{2}(x)$ are arbitrary integers which may vary from point to point. It is clear that one can not choose suitable $N_{i}(x)$($i=1,2$) so that $\widetilde{\phi}(x)$ is continuous on $R^{4}$.

In this paper, we consider the $SU(2)$ non-Abelian gauge theory for simplicity. We study the continuity of the gauge condition $n\cdot \partial n\cdot A=0$ in following texts.

\section{Continuity of the Gauge Condition $n\cdot \partial n\cdot A=0$}
\label{continuity}

In this section we consider the continuity of generator of the Wilson line
\begin{equation}
\label{WL}
W(x_{0},x_{1},x_{2},x_{3})\equiv\mathcal{P}\exp (\mathrm{i}g\int_{0}^{x_{3}}\ud z  n\cdot A(x_{0},x_{1},x_{2},z)).
\end{equation}
For a unitary matrix $U(x)=\exp(i\phi(x))$, we call the Hermitian matrix $\phi(x)$ as the generator of $U(x)$. One should not confuse it with infinitesimal generators of Lie group.

$W(x)$ is an element of $SU(2)$ group and can be written as
\begin{equation}
W(x)=e^{i\theta\vec{l}\cdot \vec{\sigma}(x)}=\cos(\theta)(x)I+\mathrm{i}\vec{\sigma}\cdot \vec{l}\sin(\theta)(x),
\end{equation}
where $\theta$ represents an arbitrary real number, $I$ represents the unit matrix in color space, $\vec{\sigma}$ represents Pauli matrixes, $\vec{l}$ represents an arbitrary unit vector.  We emphasize that $\vec{l}$ may not be well defined for $\sin(\theta)=0$. Generator of $W(x)$ can be written as $\theta\vec{l}\cdot \vec{\sigma}(x)$, which may be singular for $\sin(\theta)=0,\theta\ne 0$.

The Wilson line $W(x)$ is differentiable on $R^{4}$ given that the gauge potential $n\cdot A(x)$ is differentiable on $R^{4}$. Derivatives of $W(x)$ are continuous on $R^{4}$ given that derivatives of $n\cdot A(x)$ are continuous on $R^{4}$. However, as analysed above, generator of the Wilson line is not necessary to be continuous on $R^{4}$.

We will prove that one can always choose a continuous gauge transformation $V(x)$ on $R\otimes S^{1}\otimes S^{1}\otimes S^{1}$ so that
the generator of the Wilson line
\begin{equation}
\label{WLtr}
W^{V}(x_{0},x_{1},x_{2},x_{3})\equiv\mathcal{P}\exp (\mathrm{i}g\int_{0}^{x_{3}}\ud z  n\cdot A^{V}(x_{0},x_{1},x_{2},z)).
\end{equation}
is differentiable with continuous derivatives on $R^{4}$.

\subsection{Continuity of the generator of $W(x)$ at non-zero points of $n\cdot A(x)$}

In this subsection, we consider differentiable gauge potentials of which derivatives are continuous functions on $R\otimes S^{1}\otimes S^{1}\otimes S^{1}$. We will prove that the generator of the Wilson line (\ref{WL})  is differentiable with continuous derivatives on $R^{4}$ except for zero points of $n\cdot A(x)$.

We start from the differentiable matrix
\begin{equation}
W(x)=\cos(\theta)(x)I+\mathrm{i}\vec{\sigma}\cdot \vec{l}\sin(\theta)(x),
\end{equation}
of which derivatives are continuous $R^{4}$.
 We notice that
\begin{equation}
\cos(\theta)(x)=\frac{1}{2}tr[W(x)],\quad  \sin^{2}(\theta(x))=1-\cos^{2}(\theta)(x)
\end{equation}
and conclude that $\cos(\theta)(x)$ and $\sin^{2}(\theta(x))$ are differentiable on $R^{4}$ and derivatives of them are continuous on $R^{4}$. In addition, we can always choose the sign of the vector $\vec{l}$ so that
\begin{equation}
\sin(\theta(x))=(\sin^{2}(\theta(x)))^{1/2}.
\end{equation}
We see that  $\sin(\theta(x))$ is differentiable with continuous derivatives on $R^{4}$ except for points at which $\sin(\theta(x))=0$.  We define the angle $\theta(x)$ as
\begin{equation}
\theta(x)=\arccos(\cos(\theta(x)))
\end{equation}
and conclude that  $\theta(x)$ is differentiable with continuous derivatives on $R^{4}$ except for points at which $\sin(\theta(x))=0$.

For the vector $\vec{l}(x)$, we notice that
\begin{equation}
\vec{\sigma}\cdot \vec{l}(x)=-\mathrm{i}\sin^{-1}(\theta)(U-\cos(\theta)I)(x)
\end{equation}
and conclude that the matrix $\vec{\sigma}\cdot \vec{l}(x)$ is differentiable on $R^{4}$ except for zero points of $\sin(\theta(x))$. Derivatives of $\vec{\sigma}\cdot \vec{l}(x)$ are continuous on $R^{4}$ except for these points. The generator of $W(x)$ can be written as $\theta\vec{l}\cdot \vec{\sigma}(x)$.  Thus the generator of $W(x)$ is differentiable with continuous derivatives on $R^{4}$ except for points at which $W(x)=I$.

We then consider the behaviour of $W(x)$ in the neighborhood of points at which $W(x)=I$. Without loss of generality, we denote one of these points as $x_{0}$. We have,
\begin{equation}
\sin(\theta(x_{0}))=0,\cos(\theta(x_{0}))=1.
\end{equation}
In the neighborhood of $x_{0}$, we have,
\begin{equation}
W(x_{0}+\Delta x)=I+ign\cdot A(x_{0}) n\cdot \Delta x+\ldots,
\end{equation}
where the ellipsis represents higher order terms about $\Delta x$,  $\Delta x$ is defined as
\begin{equation}
\Delta x^{\mu}=n\cdot \Delta x n^{\mu}.
\end{equation}
We then have,
\begin{equation}
\vec{l}=\frac{1}{2(tr[n\cdot A^{2}])^{1/2}}tr[\vec{\sigma} n\cdot A]+\ldots.
\end{equation}
For the case that $\vec{l}$ is continuous at $x_{0}$, we have,
\begin{equation}
\label{phasef}
vec{l}=\frac{1}{2(tr[n\cdot A^{2}])^{1/2}}tr[\vec{\sigma} n\cdot A](x_{0}),
\end{equation}
which is well defined for $n\cdot A(x_{0})\ne 0$. According to (\ref{phasef}), we see that $\vec{l}\cdot\vec{\sigma}$ is differentiable at $x_{0}$ and derivatives of  $\vec{l}\cdot\vec{\sigma} $ are continuous at $x_{0}$ given that $\vec{l}\cdot\vec{\sigma}$ is continuous at $x_{0}$ and $n\cdot A(x_{0})\ne 0$. We notice that
\begin{equation}
\sin(\theta)(x)=-\frac{i}{2}tr[W(x)\vec{l}\cdot\vec{\sigma}(x)]
\end{equation}
and conclude that  $\sin\theta$ is differentiable at $x_{0}$ and derivatives of  $\sin\theta$ are continuous at $x_{0}$ in this case. In addition, we notice that
\begin{equation}
\ud \sin(\theta)=-\frac{\cos\theta\ud\cos\theta}{\sqrt{1-\cos^{2}\theta}},\quad \ud\theta=-\frac{\ud\cos\theta}{\sqrt{1-\cos^{2}\theta}},\quad
\cos(\theta(x_{0}))=1
\end{equation}
and conclude that $\theta$ and $\theta\vec{l}\cdot\vec{\sigma}$ are differentiable at $x_{0}$ and derivatives of them are continuous at $x_{0}$ given that $\vec{l}\cdot\vec{\sigma}$ is continuous at $x_{0}$ and $n\cdot A(x_{0})\ne 0$.

We then consider the case that $\vec{l}\cdot\vec{\sigma}$ is not continuous at a point $x_{0}$ and $W(x_{0})=I$.
We have,
\begin{eqnarray}
&&\lim_{\Delta x\to 0}\frac{W(x_{0}+\Delta x)-W(x_{0})}{|\Delta x|}
\nonumber\\
&=&\lim_{\Delta x\to 0}\frac{[\cos(\theta(x_{0}+\Delta x))-\cos(\theta(x_{0}))]}{|\Delta x|}I
\nonumber\\
&&+
i\lim_{\Delta x\to 0}\frac{\sin(\theta(x_{0}+\Delta x))}{|\Delta x|}\vec{l}\cdot\vec{\sigma} (x_{0}+\Delta x).
\end{eqnarray}
For the case that $\vec{l}\cdot\vec{\sigma}$ is not continuous at $x_{0}$,  we have,
\begin{equation}
\lim_{\Delta x\to 0}\frac{\sin(\theta(x_{0}+\Delta x))}{|\Delta x|}=0,
\end{equation}
for $W(x)$ is differentiable at $x_{0}$ and derivatives of $W(x)$ are continuous at $x_{0}$. That is, $\sin\theta$ is differentiable at $x_{0}$ and $\ud \sin\theta(x_{0})=0$. We notice that $\cos\theta$ is also differentiable at $x_{0}$ as $W(x)$ is differentiable at $x_{0}$ and have
\begin{equation}
\ud\cos\theta(x_{0})=0,\quad \ud W(x_{0})=0.
\end{equation}
We notice that
\begin{equation}
\frac{\partial W(x)}{\partial n\cdot x}|_{x
=x_{0}}=ign\cdot A(x_{0})W(x_{0})=ign\cdot A(x_{0})
\end{equation}
and conclude that $n\cdot A(x_{0})=0$ given that $\vec{l}\cdot\vec{\sigma}$ is not continuous at $x_{0}$ and $W(x_{0})=I$.

According to above proofs, we see that the generator of the Wilson line (\ref{WL}) is differentiable with continuous derivatives on $R^{4}$ except for those points at which $n\cdot A=0$.

\subsection{Gauge transformation to eliminate zero points of $n\cdot A(x)$}

In this subsection, we prove that one can always choose continuous gauge transformation $V(x)$ on $R\otimes S^{1}\otimes S^{1}\otimes S^{1}$ so that $n\cdot A^{V}(x)\ne 0$ everywhere, where
\begin{equation}
n\cdot A^{V}(x)=Vn\cdot A V^{\dag} (x)+\frac{i}{g}V n\cdot\partial V^{\dag}.
\end{equation}

$n\cdot A(x)$ is continuous on the finite compact manifold $R\otimes S^{1}\otimes S^{1}\otimes S^{1}$, so are eigenvalues of $n\cdot A(x)$. Thus maximum of  eigenvalues of $n\cdot A(x)$ does exist, which is denoted as $\lambda_{\max}$. In addition $-\lambda_{\max}$ is the minimum of eigenvalues of $n\cdot A(x)$ as   $n\cdot A(x)$ is traceless.  We consider the following gauge transformation,
\begin{equation}
V(x)=\exp(ig\frac{2\pi N n\cdot x}{L_{3}}\sigma_{3}),
\end{equation}
where $N$ represents an arbitrary integer with
\begin{equation}
N>\frac{\lambda_{max}}{2\pi L_{3}}.
\end{equation}  
One can verify that $V(x)$ is a continuous gauge transformation on $R\otimes S^{1}\otimes S^{1}\otimes S^{1}$. Under the transformation $V(x)$, we have,
\begin{equation}
n\cdot A^{V}(x)\equiv Vn\cdot A V^{\dag} (x)+\frac{i}{g}V n\cdot\partial V^{\dag}=Vn\cdot A V^{\dag} (x)+\frac{2\pi N}{L_{3}}\sigma_{3},
\end{equation}
\begin{eqnarray}
tr[\sigma_{3}n\cdot A^{V}(x)]&=&tr[Vn\cdot A V^{\dag} (x)\sigma_{3}]+\frac{4\pi N}{L_{3}}
\nonumber\\
&=&tr[n\cdot A  (x)\sigma_{3}]+\frac{4\pi N}{L_{3}}
\nonumber\\
&\ge& -2\lambda_{max}+\frac{4\pi N}{L_{3}}
\nonumber\\
&>& 0.
\end{eqnarray}
We see that one can always choose the continuous gauge transformation $V(x)$ on $R\otimes S^{1}\otimes S^{1}\otimes S^{1}$, so that $n\cdot A^{V}(x)\ne 0$ everywhere.

It is interesting to consider the limit $T\to \infty$, where $T$ is the upper bound of the time $t$ as exhibited in (\ref{timeupb}). The proof in this subsection relies on the fact that the manifold $R\otimes S^{1}\otimes S^{1}\otimes S^{1}$ is compact.  We consider the case that gauge potentials are convergent in the limit $T\to \infty$, that is,
\begin{equation}
\label{boundc}
\lim_{t\to -\infty}A^{\mu}(x)=C_{1}^{\mu}(\vec{x}),\quad \lim_{t\to \infty}A^{\mu}(x)=C_{2}^{\mu}(\vec{x}),
\end{equation}
where $C_{1}^{\mu}(\vec{x})$ and $C_{2}^{\mu}(\vec{x})$ represent arbitrary depreciable matrixes with continuous derivatives on the sub-manifold  $(S_{1})^{3}$. It is reasonable to believe that the proof in this subsection works given that eigenvalues of $C_{1}(\vec{x})$ and $C_{2}(\vec{x})$ are continuous and bounded function on the sub-manifold  $(S_{1})^{3}$.

According to proofs in this and last sections, we see that one can always choose continuous gauge transformation $V(x)$ on $R\otimes S^{1}\otimes S^{1}\otimes S^{1}$  so that the generator of the Wilson line (\ref{WLtr}) is differentiable with continuous derivatives on $R^{4}$ given that gauge potentials $A^{\mu}(x)$ are differentiable with continuous derivatives on $R\otimes S^{1}\otimes S^{1}\otimes S^{1}$.

In \cite{zhou:2016gb}, we proved that $n\cdot \partial n\cdot A$ is a continuous gauge on $R\otimes S^{1}\otimes S^{1}\otimes S^{1}$  given that the generator of the Wilson line (\ref{WL}) is continuous on $R^{4}$, where $n^{\mu}$ represents directional vectors along $x_{i}$-axis($i=1,2,3$). According to such result and the result in this section, we see that  $n\cdot \partial n\cdot A$ is a continuous gauge on the compact manifold $R\otimes S^{1}\otimes S^{1}\otimes S^{1}$ for $SU(2)$ gauge theory given that gauge potentials are differentiable with continuous derivatives on the manifold.

\section{Conclusions and Discussions}
\label{conc}

For $SU(2)$ gauge theory on the manifold $R\otimes S^{1}\otimes S^{1}\otimes S^{1}$ with finite length along every direction, we present the proof of the continuity of the gauge condition $n\cdot \partial n\cdot A=0$ given that gauge potentials are differentiable with continuous derivatives on the manifold. Given suitable boundary conditions of gauge potentials,  it is believed that the the continuity of the gauge condition does hold  even for the length of the manifold along the time direction tends to $\infty$. Compactness of the manifold considered here plays an important role in the proof of the continuity of the gauge condition $n\cdot \partial n\cdot A=0$.

As displayed in above sections, differentiability of gauge potentials is essential for the proof of  the continuity of the gauge condition $n\cdot \partial n\cdot A=0$. However, such differentiability may be destroyed by gauge transformations $U(x)$ which satisfy the equation $n\cdot \partial n\cdot A^{U}=0$. In fact, although the Wilson line (\ref{WLtr}) is differentiable, derivatives of the Wilson line (\ref{WLtr}) are not necessary  to be differentiable. Transformations of gauge potentials involving derivatives of elements in $SU(2)$ group. It is possible that gauge potentials are no longer differentiable after the gauge transformations $U(x)$. If gauge potentials are analytic on the manifold considered here, however, then it is reasonable to believe that analyticities of gauge potentials are not affected by these gauge transformations.

For the case that the length of the manifold along the time direction tends to $\infty$, suitable boundary conditions at $t\to\pm\infty$ are necessary to guarantee the continuity of the gauge condition considered here. In fact, special boundary conditions at $t\to\pm\infty$ are necessary for the vanishing of surface terms in perturbative theory. Thus we assume that the boundary condition (\ref{boundc}) does hold for quantities one concerning.

\section*{Acknowledgments}
G. L. Zhou thanks Doctor Hua-Ze Zhu for helpful discussions and important suggestions on the manuscript. The work of G. L. Zhou is supported by The National Nature Science Foundation of China under Grant No. 11647022 and The Scientific Research Foundation for the Doctoral  Program of Xi'an University of Science and Technology under Grant No. 6310116055 and The Scientific Fostering Foundation of Xi'an University of Science and Technology under Grant No. 201709. The work of Z. X. Yan is supported by  The Department of Shanxi Province Natural Science Foundation of China under Grant No.2015JM1027.

\bibliography{continuity}

\end{document}